\documentclass[preprint,3p]{elsarticle}
\usepackage[toc,page,title,titletoc,header]{appendix}
\usepackage{amsmath}
\usepackage{amssymb}
\usepackage{amsthm}
\usepackage{epsfig}
\usepackage{indentfirst}
\usepackage{doi}
\biboptions{sort&compress}
\usepackage{graphicx}
\usepackage{color}
\usepackage{mathrsfs}
\usepackage{subfigure}

\renewcommand{\vec}[1]{\mathbf{#1}}
\newcommand{\mb}[1]{\boldsymbol{#1}}

\newcommand{\nn}{\nonumber}
\newcommand{\eps}{\epsilon}
\newcommand{\rd }{\mathrm{d}}
\numberwithin{equation}{section}

\begin{document}

\begin{frontmatter}

\title{A  Level Set Method for the Simulation of  Moving Contact Lines in Three Dimensions}

\author[1]{Quan Zhao}
\address[1]{Department of Mathematics, National University of Singapore, Singapore, 119076}
\ead{matzq@nus.edu.sg}
\author[2]{Shixin Xu}
\address[2]{Zu Chongzhi Center for Mathematics and Computational Sciences,  Duke Kunshan University, 8 Duke Ave, Kunshan, Jiangsu, China}
\ead{shixin.xu@dukekunshan.edu.cn}

\author[1]{Weiqing Ren\corref{cor}}
\ead{matrw@nus.edu.sg}
\cortext[cor]{Corresponding author.}


\begin{abstract}
We propose an efficient numerical method for the simulation of multi-phase flows with moving contact lines in three dimensions. The mathematical model consists of the incompressible Navier-Stokes equations for the two immiscible fluids with the standard interface conditions, the Navier slip condition along the solid wall, and a contact angle condition which relates the dynamic contact angle to the normal velocity of the contact line (Ren et al. (2010) \cite{Ren10}). 
In the numerical method, the governing equations for the fluid dynamics
are coupled with an advection equation for a level-set function. The latter
models the dynamics of the fluid interface. Following the standard practice, the interface conditions are taken into account by introducing a singular force on the interface in the momentum equation. This results in a single set of governing equations in the whole fluid domain. Similar to the treatment of the interface conditions, 
the contact angle condition is imposed by introducing a singular force, 
which acts in the normal direction of the contact line,
into the Navier slip condition. The new boundary condition, which unifies the Navier slip condition and the contact angle condition, is imposed along the solid wall. The model is solved using the finite difference method. Numerical results, including a convergence study, are presented 
for the spreading of a droplet on both homogeneous and inhomogeneous 
solid walls, as well as   
the dynamics of a droplet on an inclined plate under gravity. 
\end{abstract}




\begin{keyword}
 Level set method \sep Multi-phase flow \sep Moving contact line 
\sep Dynamic contact angle
\sep Navier boundary condition 
\end{keyword}

\end{frontmatter}

\section{Introduction}

When two immiscible fluids are placed on the solid substrate, a moving contact
line (MCL) forms at the intersection of the fluid interface and the solid wall.
It is well-known that the classical hydrodynamics, such as the Navier-Stokes equation with the no-slip boundary condition, predicts a singularity 
for the viscous stress at the contact line. This singularity  results
 in a unphysical divergence of the rate of energy dissipation \cite{Huh71}. 
Much effort has been devoted to removing this singularity thus regularizing the model 
\cite{Dussan79,Bertozzi98,deGennes85,Jacqmin00,Kalliadasis94,Qian03,Qian06}. 
In most of these modified models, either slip or diffusion is postulated 
to occur near the MCL. In the former case, the slip is 
usually modeled by the Navier boundary condition, 
in which the shear stress is assumed to be proportional  to the slip velocity;
in the later case, a diffuse interface is employed to address the difficulty caused by the contact line singularity.

Based on molecular dynamics simulations and thermodynamics principles, 
Ren et al. proposed a set of boundary conditions for the MCL problem  
\cite {Ren07,Ren10}.
Besides the Navier boundary condition for the slip velocity, 
a condition for the contact angle is introduced at the MCL. 
When the system is at static, the contact angle of the interface
satisfied the Young's equation and 
is determined by the three surface tension coefficients.
However, when the contact line moves, the contact angle 
deviates from the static contact angle and the deviation depends 
on the contact line velocity. 
The condition for this dynamic contact angle proposed in \cite {Ren07,Ren10}
states that the unbalanced Young stress, i.e. 
the stress arising from the deviation of the dynamic contact angle, 
is balanced by the friction force at the contact line. 
The later  is proportional to
the normal velocity of the contact line. 
In this paper, we will use this model to simulate the contact line dynamics 
in three dimensions (3d). 

One of the main challenges in developing the 
numerical method is how to impose the dynamic contact angle condition.
In an earlier work \cite{Ren11, Xu14}, 
it was proposed to impose this condition through 
a singular force at the contact line. This approach is simple and 
was demonstrated to be effective in the simulation of MCLs 
in two spatial dimensions. 
In this work, we extend the method to three dimensional problems.

A number of numerical methods have been proposed for the simulation 
of multi-phase flows with MCLs, for example, in Refs. \cite{Renardy01, Qian03, Huang04, Spelt05, Afkhami08, Gerbeau2009, Yokoi09, Yue10,  Lai10, Li10, Ren11, Gao14, Xu14, Xu16, Zhang16, Solomenko17, Reusken2017, Zhao19, Zhao20ewod, zhang2020level}; 
more can be found in the review paper \cite{Sui14}. These methods generally apply different approaches to represent the fluid interface and/or different contact line conditions as
well as their numerical implementations. For example, the volume of fluid method was used to deal with the moving interface in \cite{Renardy01, Afkhami08}, and the contact angle condition was imposed on the gradient of the volume fraction function at the contact line. In Ref. \cite{Gao14}, Gao and Wang considered the phase field approach and they solved the Naiver-Stokes equations with the generalized Navier boundary condition by a splitting method based on pressure Poisson equation. In Ref.~\cite{Solomenko17}, the level set method was used to capture the fluid interface, and a dynamic contact angle model was proposed for the contact line motion dominated either by viscous effects or inertial effects. The contact angle model is then used in the reinitialization of the level set function. Recently Zhang and Yue developed a level set method in finite element framework for the 2d MCL problem \cite{zhang2020level}. In the front-tracking method, the interface is represented by a set of 
markers, and the contact line position is
updated according to either the fluid velocity at the contact line or the contact angle \cite{Huang04,Muradoglu10,Lai10, Zhang16}. By introducing the curvature as a new variable, Zhao and Ren recently proposed an energy-stable finite element method for the sharp-interface model of MCLs \cite{Zhao19, Zhao20ewod}. The dynamic contact angle condition is formulated as a time-dependent Robin type of boundary condition so that it can be naturally imposed on the curvature formulation. 
Besides, in Refs.~\cite{Reusken2017, Gerbeau2009}, the contact angle condition is absorbed into the weak form of the Navier-Stokes momentum equation by eliminating the curvature term using the surface divergence formula.

In the current work, we propose a level set method for the 3d simulation of 
MCLs. In the level set method,  the governing equations for the fluid dynamics 
are coupled with an advection equation for the level set function. 
The interface is represented by the zero-level set 
of the level set function, and the interface conditions (the jump condition 
for the normal stress and the continuity condition for the shear stress) are
taken into account in the momentum equation by adding a singular force 
accounting for the surface tension effect of the interface.
Similar idea is employed here to deal with the contact angle condition. 
Specifically, the contact angle condition is imposed by introducing a 
forcing term of the unbalanced Young stress to the Navier slip condition. 
The force depends on the dynamic contact angle and is concentrated 
at the contact line. This effectively imposes the condition for the 
dynamic contact angle. This method has been used for the simulation of
MCLs in two dimensions \cite{Ren11, Xu14}.
There, the combination of the dynamic contact angle condition 
and the Navier slip condition is straightforward since the velocity 
of the contact line aligns with the slip velocity of the fluid on the wall. 
In the 3d case, however, caution needs to be taken as the unbalanced Young
stress is only acting in the normal direction of the contact line.

The level set method for the simulation of MCLs presents another difficulty.
It is well-known that the level set function needs to be re-initialized 
from time to time so that it stays close to the signed distance function. 
This is done by solving
a Hamilton-Jacobi type of equation. When the problem involves a moving 
contact line, one needs to specify a proper boundary condition for the 
re-initialization equation. This is non-trivial, especially for 3d problems. 
In this work, we use a boundary condition which specifies the angle of the
iso-surfaces of the level set function on the wall. The angle is obtained
by the normal extension of the contact angle of the fluid interface along the
wall.

This paper is organized as follows. The contact line model, including the governing equations and the boundary conditions, 
is given in section \ref{sec:model}. Subsequently, we present the numerical 
method in section \ref{sec:method}, including the unified model 
and the level set method (section \ref{sec:lsm}) 
and the numerical discretization (section \ref{sec:num}).
Numerical examples, including the spreading of a droplet
on a homogeneous solid surface and a chemically pattern surface,  
as well as the dynamics of a droplet sliding along an
inclined plane under gravity, are presented in  section \ref{sec:nur}.
The paper is concluded in section \ref{sec:con}.

\section{Problem setup}\label{sec:model}

\begin{figure}[!t]
\centering
\includegraphics[width=0.9\textwidth]{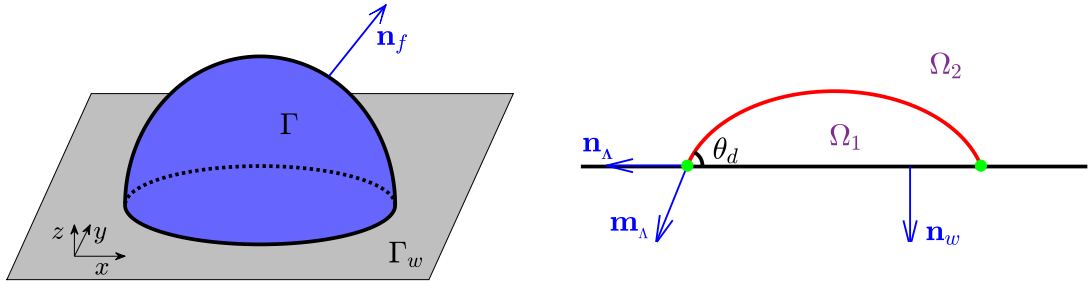}
\caption{Left panel: a schematic illustration of a droplet on a solid wall, where $\Gamma$ represents the fluid/fluid interface and $\Gamma_w$ the solid wall interface (shaded in gray) respectively. Right panel: configuration of the dynamic contact angle $\theta_d =\cos^{-1}
(\vec n_{_\Lambda}\cdot\vec m_{_\Lambda})$.
}
\label{fig:model}
\end{figure}

Without loss of generality, we consider the evolution of a droplet on a solid
substrate. A schematic setup of the system is shown in Fig.~\ref{fig:model}.  
The fluid inside and outside of the droplet are denoted by fluid 1 and 
fluid 2, respectively. The fluids are confined in the three-dimensional domain 
$\Omega=\Omega_1\cup \Omega_2$, where $\Omega_i$, $i=1, 2$,
is the region occupied by fluid $i$. The domain is bounded below by 
the solid wall $\Gamma_w$.
The interface between the two fluids (i.e. the surface of the droplet), 
denoted by $\Gamma$, intersects with the solid wall 
at the contact line $\Lambda$. The dynamics of the fluids is modeled
by the incompressible Navier-Stokes equations:
\begin{subequations}\label{ns}
\begin{align}
\rho_i\left(\partial_t\vec u+(\vec u\cdot\nabla)\vec u\right)&=\nabla p+\nabla\cdot\mb{\tau}_{d}+\rho_i\vec G, \\
\nabla\cdot \vec u&=0,
\end{align}
\end{subequations}
in $\Omega_i$ where $i=1$, 2. In the equations, $\rho_i$ is the density 
of fluid $i$, $\vec u$ is the velocity field, $p$ is the pressure, 
$\vec G$ is the body force density, and $\mb{\tau}_d$ is the viscous stress:
\begin{align}\label{navierbd}
\mb{\tau}_d=\mu_i(\nabla\vec u+(\nabla\vec u)^{T})~,
\end{align}
where $\mu_i$ is the viscosity of fluid $i$.
Across the fluid interface $\Gamma$, the velocity and the shear stress are continuous, whereas the normal stress has a jump, which is balanced by the curvature force:
\begin{align} \label{eq:interface}
[-p\mathbf{I}+\mb{\tau}_d]_1^2\cdot \mathbf{n}_f=\gamma\kappa\mathbf{n}_f, 
\end{align}
where $[\, \cdot\, ]_1^2$ denotes the jump of the physical quantity inside the bracket across the interface from fluid 1 to fluid 2, $\mathbf{I}\in\mathbb{R}^{3\times3}$ is the identity matrix, $\mathbf{n}_f$ is the unit normal vector of $\Gamma$ pointing to fluid 2, $\gamma$ and $\kappa$ are the interfacial tension and 
the mean curvature of the fluid interface, respectively.
The fluid interface evolves according to the fluid velocity,
\begin{align}
\label{kine}
\dot{\mathbf{x}}_\Gamma = \left.\mathbf{u}\right|_{_\Gamma}.
\end{align}

We use the periodic boundary condition in the $x$ and $y$ directions, i.e. the directions parallel to the wall. On the solid wall, the in-plane velocity satisfies the Navier slip condition, which states that the slip velocity is proportional to the shear stress, and the normal velocity satisfies the no-penetration condition. In vector form, these two conditions read
\begin{align}\label{slipbd}
-\beta_i \mathbf{u} =\mathbf{P}_w\cdot\mb{\tau}_d\cdot\mathbf{n}_w~,
\end{align}
where we have assumed the wall is at static, $\beta_i$ is the friction coefficient of fluid $i$ at the wall, $\mathbf{n}_w=(0, 0, -1)^T$ is the unit vector normal to the wall pointing away from the fluids, and $\mathbf{P}_w=\mathbf{I}-\mathbf{n}_w\otimes\mathbf{n}_w$ is 
projection operator onto the wall. 

At the moving contact line, we define the dynamic contact angle
as $\theta_d = \cos^{-1}(\vec n_{_\Lambda}\cdot\vec m_{_\Lambda})$,
where $\vec n_{_\Lambda}$ 
is the conormal vector of the interface between the droplet and the 
wall, and $\vec m_{_\Lambda}$ is the conormal vector of 
the fluid interface. The dynamic 
contact angle satisfies the condition \cite {Ren07,Ren10}:
\begin{align}\label{clbd}
-\beta^*\left(\vec u\cdot\vec n_{_\Lambda}\right)=\gamma \left( \cos\theta_d -\cos\theta_Y\right),
\end{align}
where $\beta^*$ is the friction coefficient at the contact line,  
$\mathbf{u}\cdot\mathbf{n}_{_\Lambda}$ is the normal speed 
of the contact line, and $\theta_Y$ is the equilibrium contact angle satisfying the Young's equations \cite{Young1805}: $\gamma_{2}-\gamma_1=\gamma\cos\theta_Y$ with $\gamma_i$, $i=1,2$, being the surface tension coefficients between fluid $i$ 
and the solid substrate. The boundary condition \eqref{clbd} is the balance of
the friction force at the contact line with the stress 
arising from the deviation of the dynamic contact angle 
from the equilibrium angle.

The governing equations in \eqref{ns} together with 
the interface and boundary conditions 
in \eqref{eq:interface}--
\eqref{clbd}
form the complete model for the evolution of the droplet 
on the solid substrate.

To make the model dimensionless, we rescale the length, time, velocity, 
pressure and the body force as
\begin{align*}
&\hat{\mathbf{x}}:=\frac{\mathbf{x}}{L}, \quad \hat{t}:=\frac{Ut}{L}, 
\quad\hat{\vec u}:=\frac{\vec u}{U},\quad \hat{p}:=\frac{p}{\rho_1\,U^2},\quad \hat{\mathbf{G}}:=\frac{ \mathbf{G}}{g},
\end{align*}
where $L$, $U$ and $g$ is the characteristic length, 
the characteristics speed and the gravitational constant, respectively.
We define the Reynolds number $Re$, the Capillary number $Ca$, 
the slip length $l_s$, the Weber number $We$ and the Bond number as
\begin{equation}
Re = \frac{\rho_1UL}{\mu_1},\quad Ca = \frac{\mu_1U}{\gamma}, 
\quad l_s = \frac{\mu_1}{\beta_1L},\quad We = Re\cdot Ca,
\quad  Bo= \frac{\rho_1 g L^2}{\gamma}.\nn
\end{equation}
Furthermore, we denote the density ratio, the viscosity ratio, 
the ratio of the two 
single-phase friction coefficients, and the ratio of the contact line 
friction coefficient to the viscosity as
\begin{align*}
&\lambda_{\rho}=\frac{\rho_2}{\rho_1},\quad \lambda_{\mu}=\frac{\mu_2}{\mu_1},\quad \lambda_{\beta} = \frac{\beta_2}{\beta_1},\quad \lambda_{\beta^*} = \frac{\beta^*}{\mu_1}.
\end{align*}
Then the dimensionless equations for the fluid dynamics 
in $\Omega_i\ (i=1,2)$ read
 \begin{subequations} \label{eqn:dmodel}
\begin{align}
\label{eq:dm1}
 \rho\left(\partial_t\vec u
 +(\mathbf{u}\cdot\nabla)\mathbf{u}\right) &=
 \frac{1}{Re}\nabla\cdot\mb{\tau}_d -\nabla p
 +\frac{1}{We}Bo\,\rho\,\mathbf{G},\\
 \nabla\cdot\mathbf{u} & =0,
 \label{eq:dm2}
 \end{align} 
 \end{subequations}
where $\tau_d= \mu (\nabla\vec u+(\nabla\vec u)^{T})$, 
 $\rho(\vec x, t)=\chi_{_{\Omega_1(t)}} + \lambda_{\rho}\,\chi_{_{\Omega_2(t)}}$,
$\mu(\vec x, t) =\chi_{_{\Omega_1(t)}} + \lambda_{\mu}\,\chi_{_{\Omega_2(t)}}$,
and $\chi_E$ denotes the characteristic function of the set $E$.
For ease of presentation, we have dropped the overhead hats
on the dimensionless variables.
The above equations are supplemented with the following conditions: 
\begin{itemize}
\item [(i)] The interface conditions on $\Gamma(t)$
\begin{align}
\label{eq:dbd1}
\left[p\,\mathbf{I} - \frac{1}{Re}\mb{\tau}_d\right]_1^2 \cdot\mathbf{n}_f&=\frac{1}{We}\kappa\,\mathbf{n}_f,  \vspace{0.15cm} \\
\dot{\vec x}_\Gamma & = \vec u |_{_\Gamma}. 
\end{align}
\item [(ii)] The boundary condition on the wall $\Gamma_w$
\begin{align}
\label{eq:dbd2}
-\beta\,\vec u = l_s\,(\mathbf{P}_w\cdot\mb{\tau}_d\cdot\mathbf{n}_w)~,
\end{align}
where $\beta(\vec x, t)=\chi_{_{\Gamma_{w,1}}} 
+ \lambda_{_\beta}\chi_{_{\Gamma_{w,2}}}$ and 
$\Gamma_{w, i}$ ($i=1,2$) is the wall in contact with fluid $i$. 
\item [(iii)] The contact angle condition at $\Lambda(t)$
\begin{align}
\label{eq:dbd3}
-\lambda_{\beta^*}\left(\vec u\cdot\vec n_{_\Lambda}\right)=\frac{1}{Ca} \left( \cos\theta_d -\cos\theta_Y\right).
\end{align}
\end{itemize}

\section{Numerical method}  \label{sec:method}

In equations \eqref{eqn:dmodel}--\eqref{eq:dbd3}, 
the fluid dynamics in $\Omega_1$ and $\Omega_2$ are modeled 
separately and coupled by the interface conditions. Next we use the
level-set approach to combine the two sets of governing equations into 
a single set of equations on the whole domain $\Omega = \Omega_1\cup\Omega_2$ 
\cite{Sussman94, Sethian1999, Xu06level, Xu03,Zhao96variational}.
The stress conditions on the fluid interface 
are imposed through an additional body force in the momentum equation. 
Similarly, we combine the dynamic contact angle
condition and the Navier slip condition into a single boundary condition
on the solid wall $\Gamma_w$. The resulting unified equations are then 
solved by the finite difference method on the whole domain.

\subsection{Unified governing equations and boundary conditions}  
\label{sec:lsm}
We use a level set function to represent and track the fluid interface. Specifically, let $\phi$ be the continuous function measuring the 
signed distance to the fluid interface,
 \begin{align}
\phi(\mathbf{x},t)=\left\{\begin{array}{ll}
 -\, \text{dist}(\mathbf{x}, \Gamma), &\mbox{for}~\mathbf{x}\in\Omega_1,\\[0.3em]
 +\, \text{dist}(\mathbf{x}, \Gamma), &\mbox{for}~\mathbf{x}\in\Omega_2.
\end{array}\right.
 \end{align}
The fluid interface at time $t$ is given by the zero-level set 
$\left\{\mathbf{x}:\phi(\mathbf{x}, t)=0\right\}$.
Using the level set function, the fluid density, viscosity and friction coefficient
can be written as
 \begin{subequations}  \label{eq:rhomubeta}
 \begin{align}
  &\rho(\mathbf{x}, t)=(1-H(\phi))+ \lambda_\rho\, H(\phi), \label{eq:rho}\\ 
 &\mu(\mathbf{x}, t) =(1-H(\phi))+ \lambda_\mu\, H(\phi), \label{eq:mu}\\
 & \beta(\mathbf{x}, t) = (1-H(\phi))+ \lambda_\beta\, H(\phi), \label{eq:beta}
\end{align} 
\end{subequations}
where $H(\phi)$ is the Heaviside function: $H = 1$ if $\phi>0$ and 0 otherwise.
The governing equations in $\Omega$ read
 \begin{subequations} \label{nsleveq}
\begin{align}
 \rho\left(\partial_t\vec u
 +(\mathbf{u}\cdot\nabla)\mathbf{u}\right) &=
 \frac{1}{Re}\nabla\cdot\mb{\tau}_d -\nabla p
 +\frac{1}{We}(\mathbf{F}+Bo\,\rho\,\mathbf{G}),\\[0.3em]
 \nabla\cdot\mathbf{u} & =0,\qquad \mathbf{x}\in\Omega
 \end{align} 
 \end{subequations}
where we have introduced a singular force
$\mathbf{F} = -\kappa \mathbf{n}_f \delta(\phi)$ 
to account for the interface condition \eqref{eq:dbd1};
$\delta(\phi)$ is the Dirac delta function, and 
\begin{align}
\label{eq:kne} 
\kappa =\nabla\cdot\left(\frac{\nabla\phi}{|\nabla\phi|}\right),
\qquad \mathbf{n}_f =\frac{\nabla\phi}{|\nabla\phi|}.
\end{align}

On the solid wall $\Gamma_w$, we combine the Navier slip condition \eqref{eq:dbd2} 
and  the contact angle condition \eqref{eq:dbd3} as follows
\begin{align}\label{nsbdlev}
-\mathbf{B} \cdot\mathbf{u}=l_s (\mathbf{P}_w\cdot\mb{\tau}_d\cdot\mathbf{n}_w)
+ \frac{1}{Ca}~\mb{\tau}_{_Y},\quad \mathbf{x}\in \Gamma_w,
 \end{align}
where  $\mathbf{B}\in\mathbb{R}^{3\times 3}$ is
 the friction coefficient tensor and 
$\mb{\tau}_{_Y}$ is the unbalanced Young stress,
 \begin{subequations}
\begin{align}
& \mathbf{B} = \beta(\mathbf{x}, t) \mathbf{I}
+ \lambda_{\beta^*} (\mathbf{n}_{_\Lambda} \otimes \mathbf{n}_{_\Lambda})
\delta(\phi),  \\
& \mb{\tau}_{_Y} = (\cos\theta_d-\cos\theta_Y)\,
\mathbf{n}_{_\Lambda}\delta(\phi).
\end{align} 
\end{subequations}
The conormal vector  $\mathbf{n}_{_\Lambda}$ 
and the dynamic contact angle $\theta_d$ can be computed 
using the level set function: 
\begin{align}
\label{eq:nlt}
\mathbf{n}_{_\Lambda} = \frac{\nabla_s \phi}{|\nabla_s\phi|},
\qquad \cos\theta_d = -\frac{\nabla\phi\cdot\mathbf{n}_w}{|\nabla\phi|}
=\frac{\partial_z\phi}{|\nabla\phi|},
\end{align}  
where $\nabla_s = (\partial_x,~\partial_y,~0)^T$ denotes 
the surface gradient along the wall. 

We note that in the sharp interface model introduced in section \ref{sec:model}, $\kappa$, $\vec n_f$ are only defined on the fluid interface, and $\vec n_{_\Lambda}$, $\theta_d$ are only defined at the contact line. In the level set method, these quantities are extended naturally in the neighbourhood of the interface\slash contact line by using the level set function $\phi$ in Eqs. \eqref{eq:kne} and \eqref{eq:nlt}. The boundary condition \eqref{nsbdlev} effectively 
imposes both the Navier slip condition and the dynamic contact angle condition. 
Indeed, away from the contact line, $\delta(\phi)$ vanishes 
so Eq. \eqref{nsbdlev} reduces to the Navier slip condition.
At the contact line, however, the two terms involving the delta function 
$\delta(\phi)$ dominate and the balance of these two terms yields
\begin{align*}
-\lambda_{\beta^*}\left(\vec u\cdot\vec n_{_\Lambda}\right)\vec n_{_\Lambda}
=\frac{1}{Ca} \left( \cos\theta_d -\cos\theta_Y\right)\vec n_{_\Lambda},
\end{align*}
which is the contact angle condition.

Finally, the level set function
evolves according to the fluid velocity:
\begin{align}\label{levup}
 \partial_t\phi+\mathbf{u}\cdot\nabla\phi=0, \quad \mathbf{x}\in \Omega\cup\Gamma_w.
\end{align}
We note that this equation governs the evolution of the level set function 
in the bulk of the fluids $\Omega$ and also on the solid wall $\Gamma_w$.
In particular, on the solid wall the equation reduces to  
\begin{equation}
 \partial_t\phi+\mathbf{u}\cdot\nabla_s\phi=0, \quad \mathbf{x}\in \Gamma_w,
\end{equation}
where we have used the no-penetration condition for the fluid velocity. 

The dynamic equations in \eqref{nsleveq} and \eqref{levup} for the fluid velocity
and the level set function, together with
the boundary condition \eqref{nsbdlev} on the solid wall, form 
a unified model for the two-phase flow with a moving contact line.
These equations are to be solved on the domain $\Omega$ by a finite difference 
method. But before we present the finite difference discretization, we discuss
the reinitialization of the level set function.

\begin{figure}[t]
\centering
\includegraphics[width=0.9\textwidth]{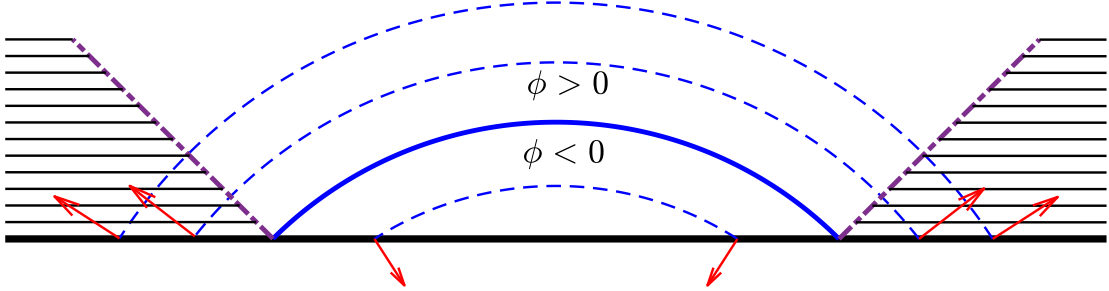}
\caption{Schematics of the level set function $\phi$, whose zero-level surface
is the fluid interface (blue solid line). The arrows are normal vectors 
to the level surfaces of $\phi$. A boundary condition is needed for the reinitialization equation 
on the wall where the normal vectors point from the wall into the fluid domain.
The level set function after the reinitialization depends on the boundary condition
in the shaded region. }
\label{2dbd}
\end{figure}

\vspace{0.5cm}
\noindent {\it Reinitialization of the level set function.}
As the interface evolves according to \eqref{levup}, the level set function
will be distorted and deviate from the signed distance function. 
To keep the level set function close to the signed distance function, it is 
a standard practice to reinitialize the function from time to time \cite{Sussman94}. This is done by solving the reinitialization equation in pseudo time $\tau$:
\begin{align}  \label{reeq}
\partial_{\tau}\phi+{\rm sgn}({\phi}_{0})(|\nabla\phi|-1)=0,\quad 
\mathbf{x}\in \Omega,
\end{align}
where ${\phi}_{0}$ is the level set function before the reinitialization,
and ${\rm sgn}(\cdot)$ is the sign function.

Most of the previous work in the level set method has considered closed 
interfaces which do not intersect with a solid wall. In the current problem,
however, due to the existence of the contact line, 
we need to specify proper boundary conditions for the reinitialization
equation \eqref{reeq} on the solid wall. 
Note that equation \eqref{reeq} is a transport equation
with velocity $\vec v_\phi = {\rm sgn}(\phi_0) \nabla\phi/|\nabla\phi|$. 
A boundary condition  is needed for this equation in regions where 
$\vec v_\phi \cdot \vec n_w <0$ on the wall.  
This is illustrated in Fig. \ref{2dbd}, where 
without loss of generality we have assumed the interface has an acute contact angle.
In this case, a boundary condition is needed on the wall outside the droplet; 
in contrast, no boundary condition can be imposed on $\phi$ on the wall 
inside the droplet where $\vec v_\phi\cdot \vec n_w >0$.

Different boundary conditions have been proposed in previous work; some 
are implicitly introduced in the numerical schemes
\cite{Spelt05,Sussman98,Sussman01,Rocca14, Griebel14}. 
In this work, we use the method proposed in Ref. \cite{Xu16}.
The reinitialization equation \eqref{reeq} is solved with the boundary condition 
\begin{equation} \label{eq:bcphi}
  \partial_z\phi =\sqrt{(\partial_x\phi)^2+(\partial_y\phi)^2}\cot\theta,
\quad \vec x=(x,y,z)\in \Gamma_{w,c},
\end{equation}
where $\Gamma_{w,c}=
\left\{\mathbf{x}\in\Gamma_w: \vec v_\phi \cdot \vec n_w <0\right\}$.
This condition requires that the iso-surfaces of $\phi$ intersect with the wall
with the angle $\theta$. The angle $\theta$ is obtained by solving
the normal extension equation on the wall 
 \begin{equation} \label{thetaeq}
\begin{cases} 
\partial_{\tau}\theta+{\rm sgn}(\phi)\,\mathbf{n}_{_\Lambda}\cdot\nabla_s\theta=0,
\quad \mathbf{x} \in \Gamma_w,\;\tau>0, \vspace{0.1cm} \\
\theta(\vec x, 0)= \arccos\left(\frac{\partial_z\phi}{|\nabla\phi|}\right).
\end{cases}
\end{equation}
where $\vec n_{_\Lambda}$ in the above equation is defined by \eqref{eq:nlt}.
By this equation, the contact angle, which is given at the contact line,
is extended to the wall following the direction normal to the contact line. In practice, $\theta$ can be first extended to the neighbourhood of the contact line
then used as the boundary condition for Eq. \eqref{reeq}; 
or Eqs. \ref{reeq} -- \ref{thetaeq} can 
be solved simultaneously. 

\subsection{Discretization}\label{sec:num}

The governing equations presented in the previous section are solved using the
 finite difference method. 
We denote by $\vec u^n$, $p^n$ and $\phi^n$ the numerical solution 
of $\vec u$, $p$ and $\phi$, respectively, at the time $t_n = n\Delta t$,
where $\Delta t$ is the time step. 
We denote by $\rho^n$, $\mu^n$ and $\beta^n$ the solution of $\rho(\vec x,t)$,
$\mu(\vec x, t)$ and $\beta(\vec x, t)$ at $t=t_n$. These quantities are 
obtained from $\phi^n$ using Eqs. \eqref{eq:rho}--\eqref{eq:beta}.
The overall procedure of the numerical method is as follows: 
at each time step $t_n$, 
\begin{itemize}
\item [1.] Update the level let function $\phi^n$ using the velocity 
$\vec u^n$ by solving Eq. \eqref{levup} for one time step 
in the bulk $\Omega$ and on the solid wall $\Gamma_w$.
This yields the solution $\tilde\phi^{n+1}$;
\item [2.] Reinitialize the level set function $\tilde\phi^{n+1}$ 
to obtain $\phi^{n+1}$ by solving Eqs. \eqref{reeq}--\eqref{thetaeq};
\item [3.] Update the fluid velocity $\vec u^n$ and pressure $p^n$ 
to obtain $\vec u^{n+1}$ and $p^{n+1}$ by solving Eq. \eqref{nsleveq} 
with the boundary condition \eqref{nsbdlev} 
on the wall and the periodic boundary condition in the $x$ and $y$ directions.
\end{itemize}

The computation domain $\Omega$ is discretized by a uniform mesh 
with mesh size $h$. 
The level set convection equation \eqref{levup} and 
the reinitialization equations in \eqref{reeq}--\eqref{thetaeq} 
are integrated using the third-order total variation diminishing 
Runge-Kutta scheme \cite{Shu88}, and the spatial derivatives in these equations
are approximated using the third-order weighted essentially non-oscillatory (WENO) 
scheme \cite{Jiang00}. 

To better conserve the volume of the droplet, we use a high-order constrained 
reinitialization technique (HCR2) proposed in Ref. \cite{Hartmann10}.
Also the local level-set technique \cite{Peng99pde, Xu16} is used 
to improve the efficiency, where we only solve the level set convection equation
and the reinitialization equations 
in small tubes around the fluid interface. 

The Navier-Stokes equations in \eqref{nsleveq} 
which involve variable density and viscosity, 
are solved using a scheme based on pressure stabilization 
\cite{Shen10,Gao14,Guermond09}. 
The discontinuous quantities (the density, viscosity and friction coefficient) 
 are smoothed out using the smoothed Heaviside function:
\begin{align}
H_{\eps}(\phi)=\left\{\begin{array}{ll}
 0,&\mbox{if} ~\phi<-\eps,\\[0.3em]
 \frac{1}{2}(1+\frac{\phi}{\eps}+\frac 1 {\pi}\sin(\frac{\pi\phi}{\eps})),& \mbox{if}~|\phi|\leqslant\eps,\\[0.3em]
 1,&\mbox{if} ~\phi>\eps,\end{array}\right.
 \end{align}
 and the singular forces are smoothed out using the mollified 
delta function:
  \begin{align} \delta_{\eps}(\phi)=\left\{\begin{array}{ll}
 0,&\mbox{if} ~|\phi|>\eps,\\[0.3em]
 \frac 1 {2\eps} (1+\cos(\frac{\pi\phi}{\eps})),& \mbox{if}~|\phi|\leqslant\eps,\end{array}\right.
 \end{align}
where we use $\eps=1.5h$ in the numerical examples.
The numerical scheme for the Navier-Stokes equations is as follows.

At the first time step when $n=0$, given $\vec u^0$ and 
the level set functions $\phi^0,~\phi^1$, we set $p^{-1}=p^0=\psi^0=0$ and obtain the velocity $\vec u^1$ by:
\begin{subequations}  \label{eq:n0}
 \begin{align}  \label{nspsm1} 
&    \frac{\frac 1 2(\rho^{n+1}+\rho^{n})\vec u^{n+1}-\rho^{n}\vec u^{n}}{\Delta t}+\rho^{n}(\vec u^{n}\cdot\nabla) \vec u^{n}+\frac{1}{2}\left(\nabla\cdot(\rho^{n}\vec u^{n})\vec u^{n}\right)\nonumber\\
&\hspace{2cm}  =\frac 1 {Re}\nabla\cdot(\mb{\tau}_d^{n+1})-\nabla(2p^{n}-p^{n-1})+\frac 1 {We} (\mathbf{F}^{n+1}+Bo\rho^{n+1}\mathbf{G}),  \quad \vec x\in \Omega, \\[0.6em]
& -\mathbf{B}^{n+1}\cdot\vec u^{n+1}=-l_s\,\mu^{n+1}\partial_z\vec u^{n+1} +\frac{1}{Ca}\mb{\tau}_Y^{n+1}, \quad \vec x\in \Gamma_w,   \label{disnsbd}
 \end{align}
\end{subequations}
and compute the pressure by:
\begin{align}\label{prepsm1}\left\{\begin{array}{lll}
&\displaystyle\Delta\psi^{n+1}=\frac {\bar{\rho}}{\Delta t}  \nabla\cdot \vec u^{n+1},&~ \mbox{in}~\Omega,\\[0.6em]
&\displaystyle\vec n_w\cdot\nabla\psi^{n+1}=0,&~\mbox{on}~\Gamma_{w},\\[0.6em]
&\displaystyle p^{n+1}=p^{n}+\psi^{n+1},&~ \mbox{in}~\Omega,
\end{array}\right.\end{align}
where we have denoted by $\vec F^{n+1}$, $\mathbf{B}^{n+1}$ and 
$\mb{\tau}_Y^{n+1}$ the respective value of $\vec F$, $\mathbf{B}$ 
and $\mb{\tau}_Y$ evaluated using $\phi^{n+1}$; 
$\mb{\tau}_d^{n+1}=\mu^{n+1}(\nabla\vec u^{n+1} + (\nabla\vec u^{n+1})^T)$, and $\bar{\rho}=\min{(1,~\lambda_{\rho})}$.

For $n\geqslant1$, given $(\vec u^{n-1},~\vec u^n,~ p^{n-1},~p^{n},~\psi^{n-1},~\psi^{n})$ and the level set functions $\phi^n,~\phi^{n+1}$, we update the velocity by:
\begin{subequations}  \label{eq:n1}
\begin{align}\label{nspsm} 
 & \rho^{n+1}\left[\frac{3\vec u^{n+1}-4\vec u^{n}+\vec u^{n-1}}{2\Delta t}+\Bigl(\left(2\vec u^n - \vec u^{n-1}\right)\cdot\nabla\Bigr) \vec u^{n}\right]+\nabla\left(p^{n}+\frac{4}{3}\psi^{n}-\frac{1}{3}\psi^{n-1}\right)\nonumber\\
 &\hspace{2cm}=\frac 1 {Re}\nabla\cdot(\mb{\tau}_d^{n+1})+\frac 1 {We} (\mathbf{F}^{n+1}+Bo\rho^{n+1}\mathbf{G}), \quad\vec x\in \Omega, \\[0.6em]
& -\mathbf{B}^{n+1}\cdot\vec u^{n+1}=-l_s\,\mu^{n+1}\partial_z\vec u^{n+1} +\frac{1}{Ca}\mb{\tau}_Y^{n+1}, \quad \vec x\in \Gamma_w, 
\label{disnsbd2} 
 \end{align}
\end{subequations}
and update the pressure by:
\begin{align}\label{prepsm}\left\{\begin{array}{lll}
&\displaystyle\Delta\psi^{n+1}=\frac {3\bar{\rho}}{2\Delta t}  \nabla\cdot \vec u^{n+1},&~ \mbox{in}~\Omega,\\[0.6em]
&\displaystyle\vec n_w\cdot\nabla\psi^{n+1}=0,&~\mbox{on}~\Gamma_{w},\\[0.6em]
&\displaystyle p^{n+1}=p^{n}+\psi^{n+1}-\mu^{n+1}\nabla\cdot\vec u^{n+1},&~ \mbox{in}~\Omega.
\end{array}\right.
\end{align}

The time discretization schemes presented above consist of a linear elliptic problem for the velocity and a Poisson equation for the pressure. To further improve the computational efficiency, we decouple the velocity components by treating $\nabla\cdot\mb{\tau}_d^{n+1}$ in \eqref{nspsm1}, \eqref{nspsm} and $\mathbf{B}^{n+1}\cdot\vec u^{n+1}$ in \eqref{disnsbd}, \eqref{disnsbd2} semi-implicitly:
\begin{align}
\nabla\cdot\mb{\tau}_d^{n+1}&\approx \mu^{n+1}\Delta\vec u^{n+1} + \nabla\mu^{n}\cdot \Bigl(\nabla\vec u^n + (\nabla\vec u^{n})^T\Bigr),\\
\mathbf{B}^{n+1}\cdot\vec u^{n+1} &\approx \mathbf{B}^{n+1}_d\cdot\vec u^{n+1} + \mathbf{B}^{n+1}_c\cdot\vec u^n,
\end{align}
where we have used the identity $\nabla\cdot\mb{\tau}_d=\mu\Delta\vec u + \nabla\mu\cdot\Bigl(\nabla\vec u + (\nabla\vec u)^T\Bigr)$ and $\mathbf{B}^{n+1}$ is decomposed into the diagonal part $\mathbf{B}_d^{n+1}$ and the off-diagonal part $\mathbf{B}_c^{n+1}$.

For the spatial discretization in Eqs. \eqref{eq:n0}--\eqref{prepsm}, 
the central difference schemes are used to 
approximate the spatial derivatives, except that 
the third-order WENO scheme is used to discretize the convection terms.
The resulting linear systems from \eqref{eq:n0} and \eqref{eq:n1}, which
have the form of discrete Helmholtz problems with varying coefficients,
are solved using the algebraic multi-grid (AMG) method or the generalized minimal residual (GMRES) method 
with the preconditioner using the periodic boundary conditions and a suitable constant coefficient.  The Fast Fourier Transform (FFT) is then applicable to the preconditioner problem as well as the Poisson equation  in \eqref{prepsm1} and \eqref{prepsm}.

\section{Numerical results}\label{sec:nur}

In this section, we first assess the accuracy and convergence of the 
numerical method  using the example of a spreading droplet on a solid substrate.
We then study the dynamic of a droplet spreading on a homogeneous 
solid substrate (example 1) and on a chemically patterned substrate 
(example 2). In the last example, we present numerical results for 
a sliding droplet on an inclined plane under the gravitational force and
compare the results with experiments.

\subsection{Convergence test}

We first study the convergence of the numerical method by considering the dynamic of a spreading droplet. The computational domain is 
$\Omega=\left\{\vec x=(x, y, z): 0\leqslant x, y\leqslant 1, 0\leqslant z\leqslant 0.5\right\}$,
where the solid surface is at $z=0$.
Initially, the droplet occupies the region $\Omega_1 = \left\{\vec x=(x, y, z): 
|\vec x-\vec x_0| \leqslant 0.25, 0\leqslant z\right\}$ with 
$\vec x_0=(0.5, 0.5, 0)$. The static contact angle of the droplet is
$\theta_Y=\pi/6$. We use
four different mesh sizes $h=1/N$ with $N=32,64,128,256$. The time step is fixed 
at $\Delta t=5\times 10^{-5}$. Other parameters are set as follows:
$\lambda_{\rho}=\lambda_{\mu}=\lambda_{\beta}=1$, 
$\lambda_{\beta^*}=1$, $Ca=0.1$, $l_s=0.1$, $Re=2$, and $Bo=0$.
To assess the numerical convergence, we compute the $L^2$ error of the numerical 
solution $u$ at time $t$  by
\begin{align}
E_N(u, t):=\left[h^3 \sum_{k=1}^{N/2}\sum_{i,j=1}^{N}
\Bigl(u_{_N}(i,j,k, t) - u_{_{2N}}(2i,2j,2k, t)\Bigr)^2 \right]^{1/2},
\end{align}
where $u_{_N}(i,j,k, t)$ is the numerical solution at $\vec x =(ih, jh, kh)$ and
time $t$ by using the mesh size $h=1/N$.

\begin{figure}[!htp]  \label{fig:interface}
\centering
\includegraphics[width=0.75\textwidth]{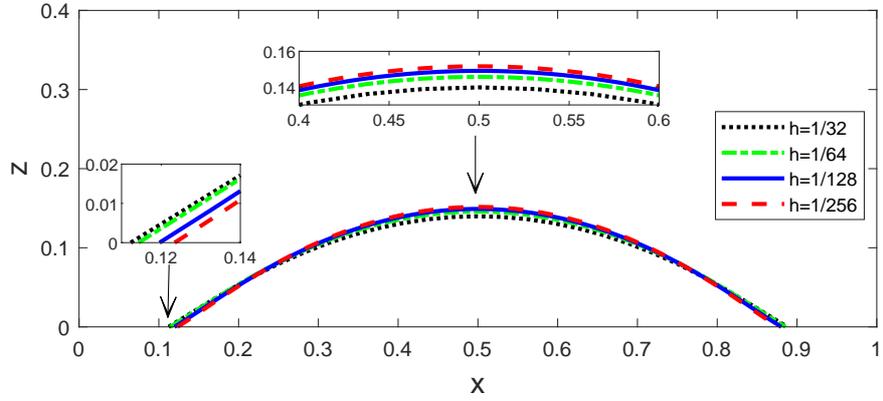}
\caption{Profile of the fluid interface at $y=0.5$ and $t=0.1$
obtained by using different mesh sizes $h$. 
The time step is $\Delta t = 5\times 10^{-5}$.}
\end{figure}

The numerical errors for the three components of the velocity 
$\vec u = (u, v, w)$ and the pressure $p$ at time $t=0.1$ 
are shown in Table \ref{tb:order0}. We observe the decrease of 
the numerical errors as the mesh is refined. The order of convergence 
reaches about one for the velocity components, but only around one half 
for the pressure. The lower convergence order for the pressure could be 
due to its singular behaviour at the moving contact line. 
Furthermore, we plot the zero level surface of $\phi$ for 
the interface at $y=0.5$ and $t=0.1$ in Fig. \ref{fig:interface}. 
The numerical convergence for the interface as the mesh is refined is observed. 

\begin{table}[!htp]
\centering
\def\temptablewidth{0.95\textwidth}
\vspace{-12pt}
\caption{Numerical errors and the convergence rate for the velocity components 
$u$, $v$ and $w$, and the pressure $p$ at time $t=0.1$ 
under different mesh sizes $h=1/N$.
the time step is fixed at $\Delta t= 5\times 10^{-5}$. }
{\rule{\temptablewidth}{1pt}}
\begin{tabular*}{\temptablewidth}{@{\extracolsep{\fill}}lllllllll}
$N$ & $E_N(u,~t)$ &order  & $E_N(v,~t)$ &order & $E_N(w,~t)$ &order  & $E_N(p,~t)$ &order  \\ \hline
32 &1.46E-2 &- & 1.46E-2 & - &1.12E-2 &-& 1.10E0 &- \\ \hline
64 &5.88E-3 &1.31 & 5.88E-3 & 1.31 &5.60E-3 &1.00& 7.80E-1 &0.50  \\ \hline
128 &2.81E-3 &1.07 & 2.81E-3 & 1.07 &3.00E-3 &0.90& 6.06E-1 &0.36 
 \end{tabular*}
{\rule{\temptablewidth}{1pt}}
\label{tb:order0}
\end{table}

\subsection{Applications}
{\bf Example 1.} Similar to the system used in the convergence test, 
we consider a droplet spreading on a solid substrate 
with the static contact angle $\theta_Y=\pi/6$. 
The initial configuration of the droplet and the computational domain are the
same as that in the convergence test. Other parameters are as follows:
$\lambda_{\rho}=\lambda_{\mu}=0.2$, $\lambda_{\beta}=\lambda_{\beta^{*}}=1$,
$Re=2$, $Ca=0.1$, $l_s=0.1$ and  $Bo=0$.
 The mesh size and the time step are $h=1/128$ and $\Delta t= 5\times 10^{-5}$,
respectively.
 
 \begin{figure}[htp]
\centering
\includegraphics[width=0.8\textwidth]{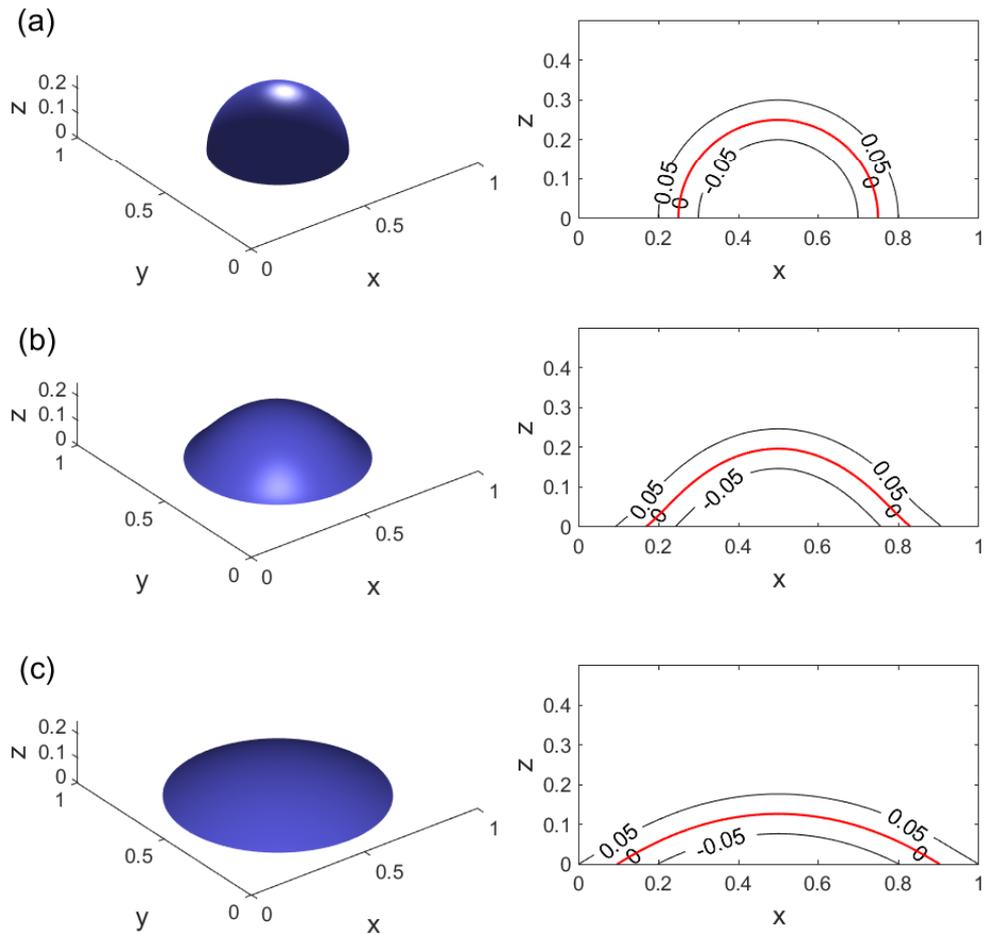}
\caption{Snapshots of the droplet profile (left panel) and the corresponding contour of the reinitialized level set function at $y=0.5$ (right panel) in the evolution of a droplet spreading towards the equilibrium at three times. (a) $t=0$; (b) $t=0.05$; (c) $t=0.25$. The red thick line denotes the fluid interface. }
\label{fig:spreadDrop}
\end{figure}

The deviation of the contact angle from the equilibrium contact angle $\theta_Y$ 
gives rise to an unbalanced Young's stress at the contact line, 
which drives the droplet towards the equilibrium.
In Fig.~\ref{fig:spreadDrop}, we present 
several snapshots of the evolving droplet, together with the corresponding 
level set function. After reaching the steady state, the droplet takes the 
shape of a spherical cap with the equilibrium contact angle $\theta_Y$. 

\begin{figure}[htp]
\centering
\includegraphics[width=0.80\textwidth]{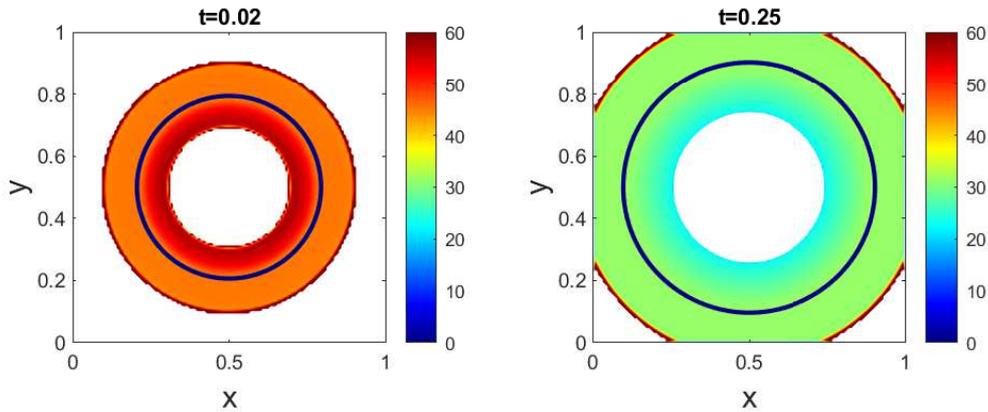}
\caption{The angle of the iso-surfaces of the reinitialized level set function
at the wall. The angle equals the contact angle extended from the contact line
outside the droplet. The black thick line represents the moving contact line. }
\label{fig:spreadAngle}
\end{figure}

In the computation, the level set function $\phi$ is reinitialized by solving 
Eq. \eqref{reeq} with the boundary condition \eqref{eq:bcphi}. 
The boundary condition,
which requires the iso-surfaces of $\phi$ meet with the wall at
the angle extended from the contact line, is imposed at the wall 
outside the droplet. In Fig.~\ref{fig:spreadAngle}, 
we show the angle of the iso-surfaces of the reinitialized level set 
function at the wall near the contact line at two different times. 
Indeed, in the region outside the droplet this angle agrees 
with the contact angle of the fluid interface.

\begin{figure}[htp]
\centering
\includegraphics[width=0.85\textwidth]{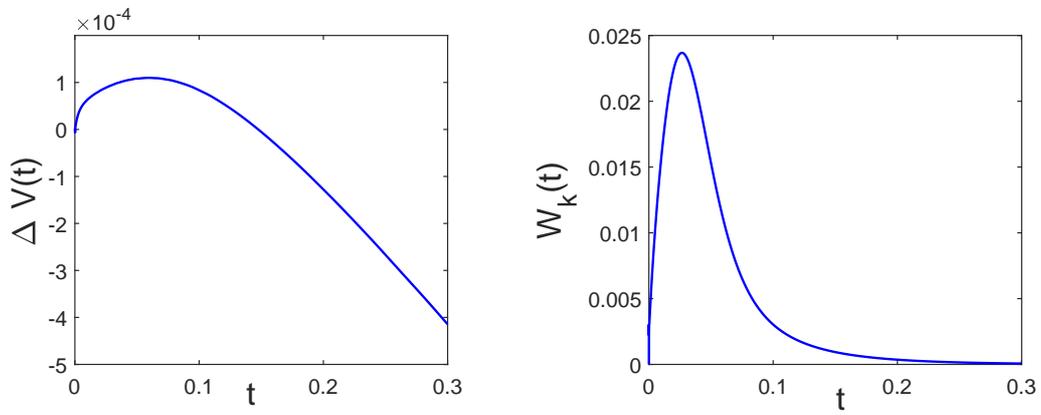}
\caption{The relative volume loss $\Delta V(t)$ (left panel) and 
the kinetic energy $W_k(t)$ (right panel) versus time, for the 
relaxing droplet in example 1.  
}
\label{fig:volumekinetic}
\end{figure}

Finally, we consider the histories of the droplet volume and its kinetic energy.
We define the relative volume loss  and the kinetic energy of the system as
\begin{align}
\Delta V(t):=\frac{V(t) - V(0)}{V(0)},\qquad 
W_k(t):=\frac{1}{2}\int_{\Omega}\rho|\vec u|^2\,\rd V,\nn
\end{align}
 where $V(t):=\int_{\Omega}\left(1-H_{\epsilon}(\phi(\vec x, t))\right)\rd V$ is the volume of the droplet in discrete form. The time history of these two quantities are shown in Fig.~\ref{fig:volumekinetic}. We observe the volume of the droplet is well conserved with relative change on the order of $10^{-4}$
during the time evolution. The kinetic energy first builds up then gradually 
decays to zero as the system evolves towards the equilibrium.

\begin{figure}[tph]
\centering
\includegraphics[width=0.85\textwidth]{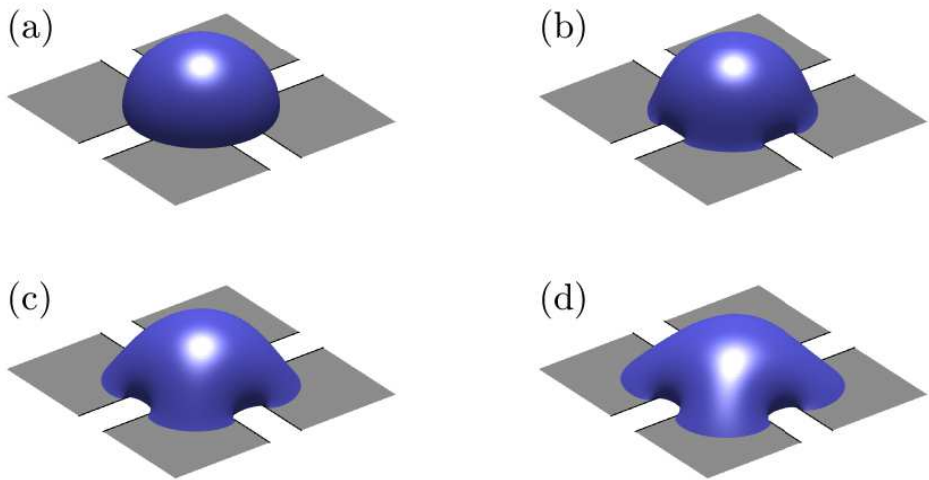}
\caption{Snapshots of the droplet spreading on a chemically patterned 
surface at several times: (a) $t=0$; (b) $t=0.05$; (c) $t=0.1$; (d) $t=0.35$. 
The static contact angle is $\theta_Y=2\pi/3$ on the two strips
in the middle of the solid surface and $\theta_Y=\pi/4$ outside the strips. }
\label{fig:ChePa}
\end{figure}

\begin{figure}[thp]
\centering
\includegraphics[width=0.85\textwidth]{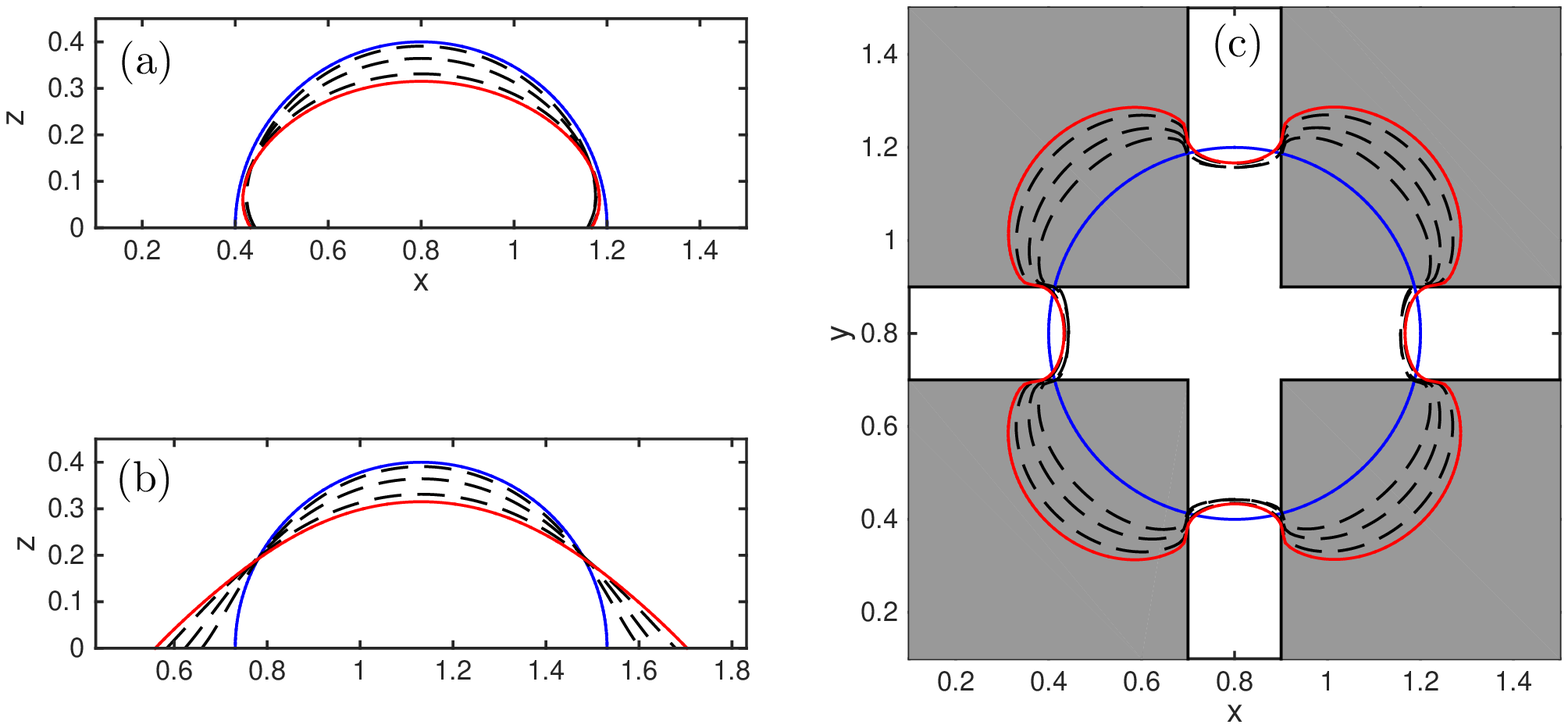}
\caption{(a) Cross-section of the interface at $y=0.8$; 
(b) Cross-section of the interface at $y=x$; 
(c) Evolution of the contact line . 
The blue curve corresponds to $t=0$, and the  
red curve corresponds to the final time $t=0.35$.}
\label{fig:ChePa1}
\end{figure}

\vspace{0.5em}
{\bf Example 2.} In this example, we investigate the dynamic of 
a droplet spreading on an inhomogeneous solid surface \cite{Yan2007lattice}.
As shown in Fig.~\ref{fig:ChePa}, the solid surface contains two narrow 
hydrophobic strips. The static contact angle of the droplet
is $\theta_Y=2\pi/3$ on the strips, and $\theta_Y=\pi/4$ outside the strips. 
The computational domain is 
$\Omega=\left\{\vec x=(x, y, z): 0\leqslant x, y\leqslant 1.6,\ 0\leqslant z\leqslant 1.2\right\}$,
where the solid surface is at $z=0$.
The two strips occupy the region $S=S_1\cup S_2$, where
$S_1 = \left\{\vec x \in \Gamma _w: |x-0.8| \leqslant 0.1\right\}$ and
$S_2 = \left\{\vec x \in \Gamma _w: |y-0.8| \leqslant 0.1\right\}$.
Initially, the droplet has a spherical shape and occupies the region 
$\Omega_1 = \left\{\vec x=(x, y, z): |\vec x-\vec x_0| \leqslant 0.4,\ 
0\leqslant z\right\}$ with $\vec x_0=(0.8, 0.8, 0)$. 
Other parameters in the system are chosen as: 
$\lambda_{\rho}=\lambda_{\mu}=0.2$, $\lambda_{\beta}=\lambda_{\beta^*}=1$,
$l_s=0.1$, $Re=2$, $Ca=0.1$ and $Bo=0$.
The mesh size is $h=0.01$, and the time step is 
$\Delta t=1\times 10^{-4}$.

The numerical results are shown in Fig.~\ref{fig:ChePa}, where we present the 
snapshots of the droplet at several times. As expected, the droplet, which 
is initially given by a hemisphere with contact angle $\pi/2$, 
contracts inwards on the hydrophobic strips and spreads outwards 
on the hydrophilic regions. Cross-section profiles of the interface 
and the contact line at different times are shown in Fig.~\ref{fig:ChePa1}.

 \begin{figure}[thp]
\centering
\includegraphics[width=0.6\textwidth]{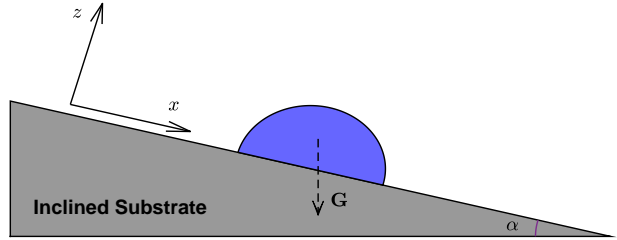}
\caption{Illustration of a droplet sliding on an inclined plane of angle 
$\alpha$ under the gravitational force $\vec G$. }
\label{fig:inclined}
\end{figure}

\vspace{0.5em}
{\bf Example 3.} 
In this example, we simulate the dynamics of a droplet on an inclined plane 
under gravitational force. 
A cross section of the system is shown in Fig.~\ref{fig:inclined}. 
The $xy$-plane of the coordinate system is aligned with the solid surface,
and the computational domain is 
$\Omega=\left\{\vec x=(x, y, z): 0\leqslant x\leqslant 4, ~0\leqslant y\leqslant 3,~0\leqslant z\leqslant 1\right\}$.
The static contact angle of the droplet is $\theta_Y=48^\circ$.
Initially, the droplet takes the shape of a spherical cap and occupies the region 
$\Omega_1 = \left\{\vec x=(x, y, z): |\vec x-\vec x_0| \leqslant 1.484,\ 
0\leqslant z\right\}$ with $\vec x_0=(2, 1.5, -0.993)$. 
Other parameters in the system are chosen as
$\lambda_\rho=1.3\times 10^{-3}$,
$\lambda_\mu=0.2$, $\lambda_\beta=1$, $\lambda_{\beta^*}=30$, $l_s=0.02$,
$Ca=2.5\times 10^{-3}$, $Re=0.85$ and $Bo=1.54$.

\begin{figure}[thp]
\centering
\includegraphics[width=0.85\textwidth]{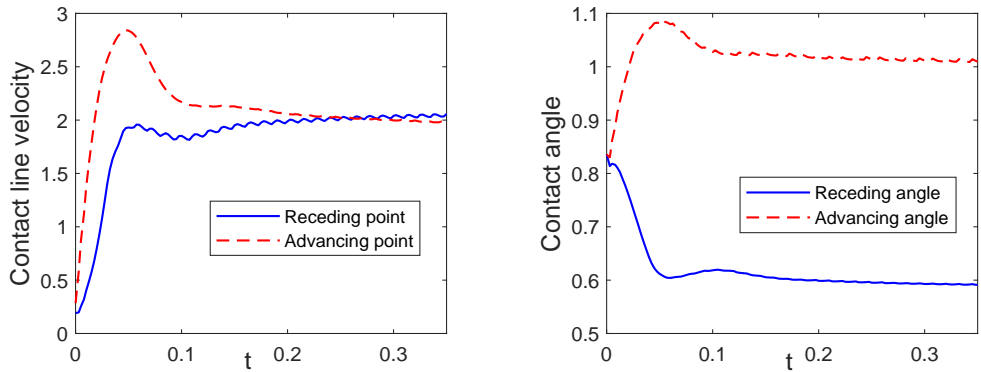}
\caption{The contact line speed (left panel) and the dynamic contact angle 
(right panel) measured at the advancing contact point $x_a$ and the
receding contact point $x_r$.
The inclination angle of the plane is $\alpha=\pi/6$.}
\label{fig:CLV}
\end{figure}

\begin{figure}[!htp]
\centering
\subfigure[]{
\includegraphics[width=0.7\textwidth]{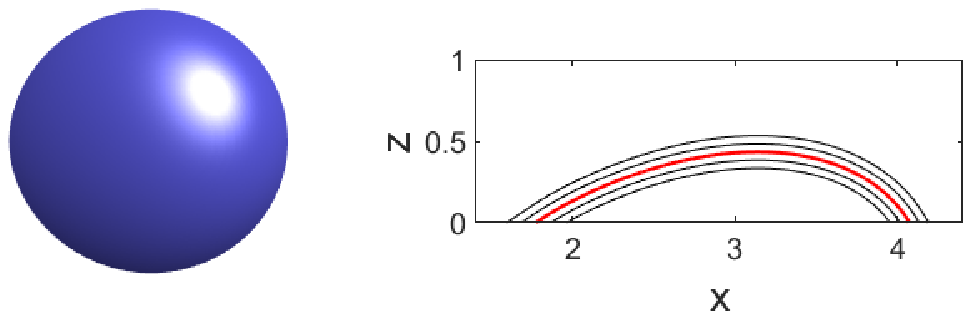}}
\subfigure[]{
\includegraphics[width=0.7\textwidth]{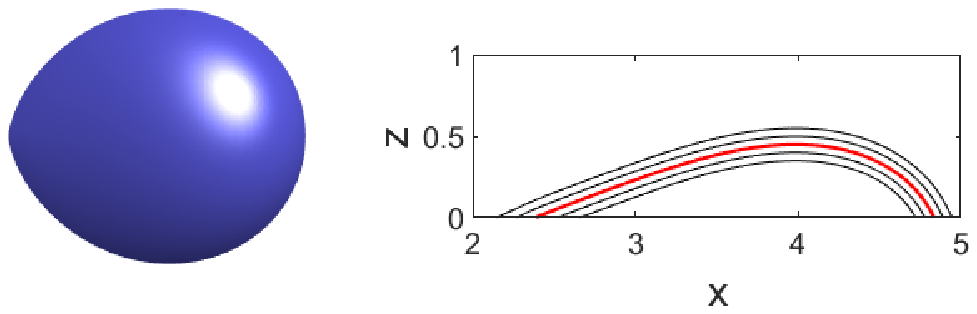}}
\subfigure[]{
\includegraphics[width=0.7\textwidth]{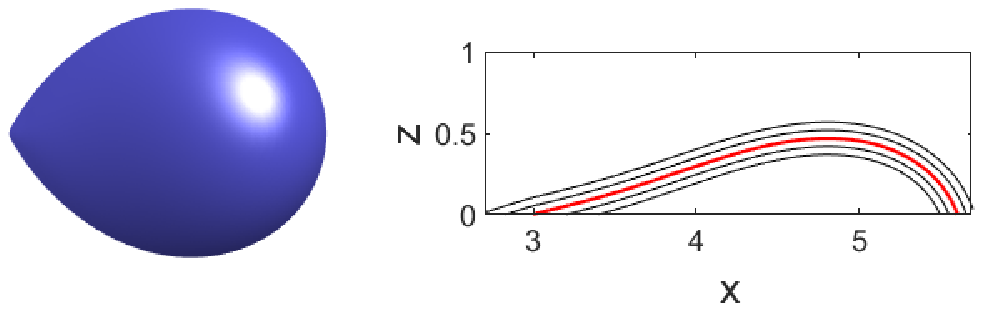}}
\caption{ Left panels: top view of the steady-state configurations 
of the sliding droplet on a substrate with different inclination angles:
 (a) $\alpha=30^\circ$; (b) $\alpha=50^\circ$; (c) $\alpha=70^\circ$.
Right panels: contours of the level set function at $y=1.5$ (the plane of 
symmetry), where the red curve is the fluid interface. 
} \label{fig:inclinedDroplet}
\end{figure}

In the numerical experiments, we use the mesh size $h = 1/40$ and the time step 
$\Delta t = 5\times 10^{-5}$. We first set the inclination angle of the plane
at $\alpha=\pi/6$. In Fig. \ref{fig:CLV} (left panel), we plot the speeds of 
the advancing contact point $x_a$ and the receding contact point $x_r$,
where $x_a = \max_{\vec x\in \Lambda} x$ and $x_r = \min_{\vec x\in \Lambda} x$.
Due to symmetry, $x_a$ and $x_r$ are located on the line $y=1.5$ on the plane.
From the figure we see that the two speeds reach about the same value 
after $t\approx 0.25$. After this time, the droplet slides along the plane 
at this steady-state speed. Also shown in the figure (right panel) are 
the dynamic contact angles measured at 
the advancing and receding contact points.
Both angles differ from the static contact angle $\theta_Y$. They are 
related to the contact line speed via the condition in Eq. \eqref{eq:dbd3}.

\begin{figure}[thp]
\centering
\includegraphics[width=0.80\textwidth]{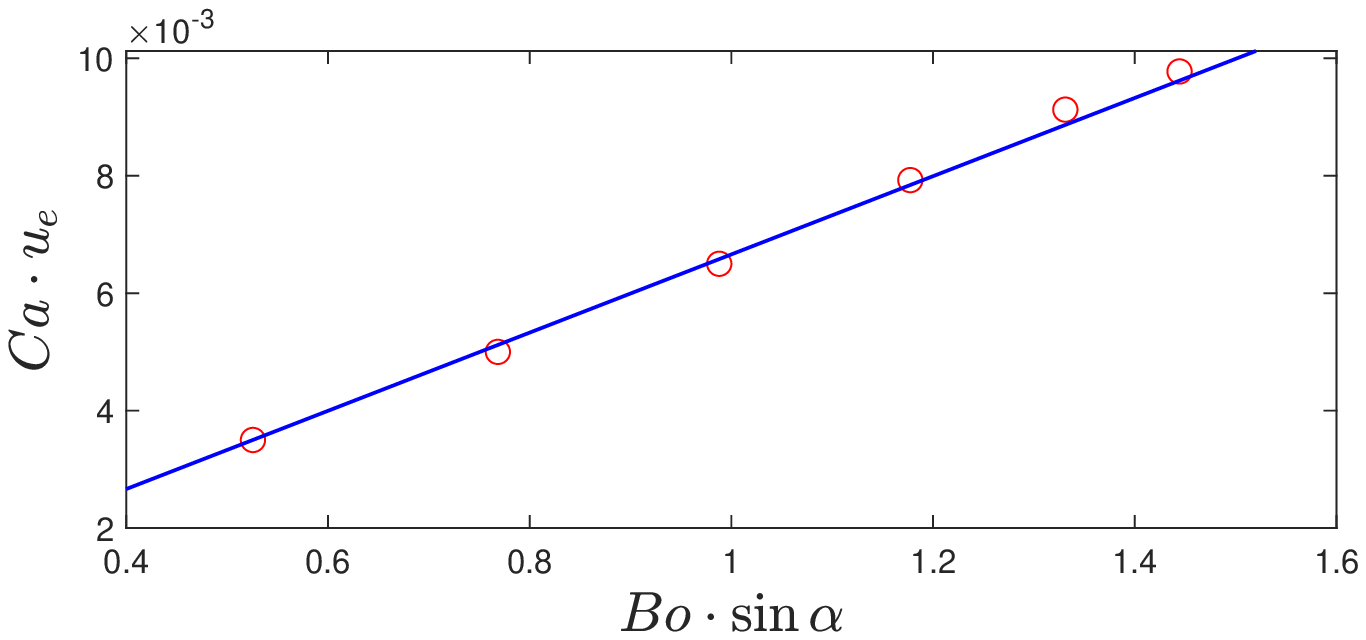}
\caption{ Relation between the steady-state speed of the droplet
$u_e$ and the inclination angle $\alpha$. The discrete circles 
are the numerical results and the blue line is the fitting function:
$Ca\cdot u_e = 0.00666Bo\cdot \sin\alpha$. }
\label{fig:BOR}
\end{figure}

We then change the inclination angle of the substrate and simulate 
the corresponding dynamics of the droplet. The profiles of the droplet 
at the steady state are presented in Fig.~\ref{fig:inclinedDroplet},
where the different configurations correspond to 
different inclination angles. 
We observe the formation of an oval shape for the sliding droplet as the
inclination angle is increased. Also with the increasing inclination angle,
the moving contact line develops a sharp corner at the tip.
In Fig. \ref{fig:BOR}, we plot the steady-state speed of the droplet
versus the sine of the inclination angle. The speed increases with the inclination
angle, and the relation is well-fitted by a linear function. 
These numerical results agree with those observed in experiments \cite{Le2005shape}. 

\section{Conclusion}\label{sec:con}

In this work, we developed an efficient numerical method for the simulation
of two-phase flows with moving contact lines in three dimensions. 
The fluid dynamics, which was modeled by the two-phase incompressible 
Navier-Stokes equations, was coupled with an advection equation 
for the level-set function. The latter models the dynamics 
of the fluid interface and the moving contact line.
The interface conditions were taken into account by introducing
a singular force in the momentum equation; similarly 
the contact angle condition was imposed by introducing a singular force in the 
Navier slip boundary condition. 
The unified model was then solved in the whole fluid domain.
Based on the pressure stabilization,  
we proposed a finite difference method for solving the Navier-Stokes 
equations with the unified boundary condition.

The reinitialization of the level set function requires a proper boundary
condition on the solid wall where the contact line moves.
We employed an angle condition which specifies the angle of 
the iso-surfaces of the level set function. The angle was obtained 
by the normal extension of the contact angle of the fluid
interface from the contact line. In practice, the reinitialization 
equation for the level set function and the extension equation for the angle
can be solved simultaneously in  time.

The performance of the numerical method was illustrated by using several examples. The numerical results for a spreading droplet with different mesh sizes
showed the convergence of the numerical method. 
The simulation of a droplet spreading on a chemically patterned 
wall demonstrated the applicability and efficiency of the numerical method for
problems with complex boundary conditions.
In addition, the simulation for the sliding droplet on inclined wall 
yielded consistent results as in the experiments.
This method enables us to study interesting physical processes 
in multi-phase flows with moving contact lines in three dimensions. 
These applications will be left to our future work.
 
\section*{Acknowledgement}
The work of Ren was partially supported by Singapore MOE AcRF grants 
(R-146-000-285-114, R-146-000-327-112) and NSFC (NO. 11871365).

\bibliographystyle{elsarticle-num}
\bibliography{thebib}

\end{document}